\documentclass{elsart}

\usepackage{amssymb}
\usepackage{graphicx}

\begin{document}

\begin{frontmatter}

\title{Superdeformed band in the $N = Z+4$ nucleus $^{40}$Ar:
A projected shell model analysis}

\author[a1]{Ying-Chun Yang,} \thanks{Present address: Shanghai Jiao Tong University Press, Shanghai 200030, P. R. China}
\author[a2]{Yan-Xin Liu,}
\author[a1,a3,a4]{Yang Sun,} \thanks{Corresponding author: sunyang@sjtu.edu.cn}
\author[a5]{Mike Guidry}

\address[a1]{Department of Physics and Astronomy, Shanghai Jiao Tong University,
Shanghai 200240, P. R. China}
\address[a2]{School of Science, Huzhou Teachers College, Huzhou, 313000, P. R. China}
\address[a3]{IFSA Collaborative Innovation Center, Shanghai Jiao Tong
University, Shanghai 200240, P. R. China}
\address[a4]{State Key Laboratory of Theoretical Physics, Institute of
Theoretical Physics, Chinese Academy of Sciences, Beijing 100190, P.
R. China}
\address[a5]{Department of Physics and Astronomy, University of
Tennessee, Knoxville, TN 37996, USA}

\begin{abstract}

It has been debated whether the experimentally-identified superdeformed 
rotational band in $^{40}$Ar [E. Ideguchi, et al., Phys. Lett. B 686 (2010) 18] 
has an axially or triaxially deformed shape.  Projected shell model calculations 
with angular-momentum-projection using an axially-deformed basis are performed 
up to high spins.  Our calculated energy levels indicate a perfect 
collective-rotor behavior for the superdeformed yrast band.  However, detailed 
analysis of the wave functions reveals that the high-spin structure is dominated 
by mixed 0-, 2-, and 4-quasiparticle configurations. The calculated electric 
quadrupole transition probabilities reproduce well the known experimental data 
and suggest a reduced, but still significant, collectivity in the high spin 
region.  The deduced triaxial deformation parameters are small throughout 
the entire band, suggesting that triaxiality is not very important for this 
superdeformed band.

\end{abstract}

\end{frontmatter}

\newcommand{\qt}{Q_{\rm t}}
\newcommand{\mpole}{G_{\rm M}}
\newcommand{\qpole}{G_{\rm Q}}

Nuclei are among the few quantum systems that can be described meaningfully in 
terms of shape. Understanding the known shapes and the search for exotic shapes 
has long been a research forefront in nuclear structure physics 
\cite{Frauendorf01}. One example is the study of superdeformed (SD) shapes and 
the associated collective rotational bands \cite{SD-review}, which have been  
identified experimentally in various mass regions of the nuclear chart 
\cite{SD-data}. It has been suggested that the high-spin behavior of an SD band 
is often strongly influenced by the high-$j$ intruder orbits because of the 
unique properties exhibited by these orbits with rotation.  For example, 
variations in the SD yrast band structure in the mass-190 region are 
characterized by competition between rotational alignment of a pair of 
quasi-protons from the $\pi 1i_{13/2}$ orbit and a pair of quasi-neutrons from 
the $\nu 1j_{15/2}$ orbit \cite{SD190}.  As another example, the rotational 
alignment of the $1g_{9/2}$ quasi-proton and quasi-neutron pairs dominates the 
high-spin behavior of  SD nuclei in the mass-60 region \cite{SD60}.  These 
examples demonstrate that by studying high-spin states of  SD bands one can gain 
microscopic insights into the role and nature of the high-$j$ intruder orbits. 
This is very useful information, particularly for those exotic mass regions for 
which the single-particle structure is less well known.

A few $N=Z$ nuclei in the mass-40 region ($^{36}$Ar \cite{Ar36a}, $^{40}$Ca 
\cite{Ideguchi01}, and $^{44}$Ti \cite{Leary00}) have been found to exhibit 
superdeformed structure. Ideguchi {\it et al.} \cite{Ideguchi10}  used the 
$^{26}$Mg($^{18}$O, 2p2n)$^{40}$Ar reaction to establish to higher spin a 
rotational band in $^{40}$Ar and to show that it has an unusually large 
transition quadrupole moment $\qt=1.4^{+0.49}_{-0.31}$ eb.  The lower-spin 
states of the $^{40}$Ar SD band were known previously along with some 
limited electric quadrupole transition information \cite{BE2DATA04}, but the 
Ideguchi {\it et al.} measurement of such a large $\qt$ value indicates 
clearly that this is a SD band, and the extension to high spins allows one 
to study excitations of the high-$j$ particles in the superdeformed well where 
rotation alignment of these particles takes place. The observed high-spin 
behavior in this $^{40}$Ar SD band clearly differs from that of the SD band in 
the $N=Z$ isotope $^{36}$Ar \cite{Ar36a,Ar36b,Long01}, suggesting that the 
addition of four neutrons to $^{36}$Ar has significant impact in the SD 
structure.   In addition, as pointed out in Ref.\ \cite{Ideguchi10}, $^{40}$Ar 
is an $N = 22$ isotone of $^{34}$Mg, which was found to have a SD ground state 
\cite{Mg34} and is one of only a few nuclei lying in the suggested ``island of 
inversion" \cite{Brown90}.

It is remarkable that quantum shell effects can stabilize a superdeformed shape 
in nuclei having such a small number of particles.  This has been interpreted 
according to two ideas: (i)~multiparticle--multihole (mp--mh) excitation from 
the sd to fp shell, and (ii)~emergence of the SD shell gaps at $N = Z = 18$, 20, 
and 22. The structure of the SD band in $^{40}$Ar was studied using cranked 
Hartree--Fock--Bogoliubov (CHFB) theory in Ref.\ \cite{Ideguchi10}. These 
calculations reproduced reasonably well the energy levels of the SD band, and 
the analysis showed that this band corresponds to a mp--mh excitation across the 
sd--fp shell gap. Electromagnetic transition properties were not discussed.  The 
CHFB is a mean-field approximation and the discussion and conclusions were 
obtained within a cranked intrinsic framework. This CHFB analysis 
\cite{Ideguchi10} suggested that a simultaneous alignment of the $f_{7/2}$ 
protons and neutrons that is central to explaining the $^{36}$Ar SD 
structure \cite{Long01} does not occur in $^{40}$Ar, that the $^{40}$Ar SD band 
is associated with an axially deformed shape, and that its triaxiality is very 
small ($\gamma\approx$ 0) for the entire angular-momentum range.

In contrast to Ref.\ \cite{Ideguchi10}, Taniguchi {\it et al.} \cite{Triaxial} 
proposed that the observed SD band in $^{40}$Ar could be interpreted as 
resulting from triaxial superdeformation. Their analysis used an 
antisymmetrized molecular dynamics plus generator coordinate method (AFT+GCM), 
which includes effects beyond those for typical mean fields.  However,  the 
AFT+GCM calculations were able to describe the observed data for energy levels 
and $B(E2)$ values at only a qualitative level, and the triaxiality found  was 
not large ($\gamma\approx 10^\circ$).  To summarize, the mean-field CHFB 
calculation \cite{Ideguchi10}, which reproduces data well,  suggests an 
axially-symmetric superdeformed shape for the observed SD band in $^{40}$Ar, 
while the AMD+GCM approach \cite{Triaxial}, which includes more many-body 
correlations but does not describe the $^{40}$Ar data as well, suggests a SD 
band built on a triaxially-deformed shape.  It would be desirable to understand 
and resolve these different interpretations for the observed SD band in 
$^{40}$Ar.

Understanding the detailed structure of $^{40}$Ar could also benefit
other studies. For example, properties of excited states in
$^{40}$Ar are of astrophysical importance because neutrino-induced
reactions on $^{40}$Ar are used to detect the solar neutrino emitted
from $^8$B in the Sun through the liquid argon time projection
chamber (LArTPC) in ICARUS (Imaging of Cosmic and Rare Underground
Signals) \cite{Rubbia98}.  It was pointed out in Ref.\
\cite{Niu14} that it is possible for the SD state in this nucleus to
accommodate an $\Lambda$ particle to form an hypernucleus.

The projected shell model (PSM) \cite{PSM} was applied to analyze the structure 
of the earliest known SD bands \cite{SD130}. In Ref.\ \cite{Long01}, some of the 
present authors applied the PSM to study the SD band in the $N=Z$ nucleus 
$^{36}$Ar \cite{Ar36a,Ar36b}, and interpreted the band disturbance around $I=12$ 
as a consequence of the simultaneous alignment of the $1 f_{7/2}$ quasi-proton 
and neutron pairs, which was later supported by other studies 
\cite{Inakura02,Poves04,Oi07,Bender03}.  In the present Letter, we analyze the 
SD band in $^{40}$Ar  using the PSM. 

The PSM \cite{PSM} is a shell model truncated in the deformed Nilsson 
single-particle basis, with pairing correlation incorporated into the basis by a 
BCS calculation for the Nilsson states.  The shell-model truncation is first 
implemented in the multi-quasiparticle (qp) basis with respect to the deformed 
BCS vacuum $\left|0\right>$, then the basis states are transformed from the 
intrinsic to the laboratory frame by angular momentum projection \cite{PSM}, and 
finally a two-body shell model Hamiltonian is diagonalized in this projected 
space.  Unlike the CHFB, the PSM goes beyond mean-field by transforming the 
basis from the intrinsic to the laboratory frame and performing configuration 
mixing.  The PSM treatment is distinguished from the AMD+GCM approach by 
explicit inclusion of mp--mh (multi-qp) configurations in the diagonalization.  
These are helpful ingredients in the investigation of the current problem and 
in resolving present conflicting interpretations of the $^{40}$Ar SD structure.

The PSM wavefunction is a superposition of
angular-momentum-projected, multi-qp states that span the shell model
space
\begin{equation}
\left|\Psi^{\sigma}_{IM}\right> = \sum _{K \kappa}
f^{\sigma}_{IK_\kappa}\,\hat P^I_{MK}\left|\Phi_\kappa \right> ,
\label{wf}
\end{equation}
where
\begin{equation}
\hat{P}^{I}_{MK} = \frac{2I+1}{8\pi^2}\int d\Omega
D^{I}_{MK}(\Omega)\hat{R}(\Omega) \nonumber
\end{equation}
is the angular momentum projector \cite{PSM}.  In Eq. (\ref{wf}),
$\left|\Phi_\kappa\right\rangle$ denotes the qp-basis, $\kappa$
labels the basis states, and $f^{\sigma}_{IK_\kappa}$ are determined
by the configuration mixing implemented by diagonalization.  In the
present work, the deformed single-particle basis is generated with a
quadrupole deformation parameter $\varepsilon_2=0.48$, which is
consistent with the experimental deformation $\beta_2\sim
0.5$ \cite{Ideguchi10} for this SD band.  The
single-particle basis is taken to be axially symmetric but 
asymmetry can enter the solutions through configuration mixing, as we shall
discuss later. Particles in three major shells
($N=1,2,3$ for both neutrons and
protons) are activated to define the valence space. 
The multi-qp basis $\left|\Phi_\kappa \right>$ (including up to 4-qp states) is
taken as
\begin{equation}
\begin{array}{rl}
\{ \left|0 \right\rangle, a^\dagger_{\nu_i} a^\dagger_{\nu_j}
\left|0 \right\rangle, a^\dagger_{\pi_i} a^\dagger_{\pi_j} \left|0
\right\rangle, a^\dagger_{\nu_i} a^\dagger_{\nu_j} a^\dagger_{\pi_k}
a^\dagger_{\pi_l} \left|0 \right\rangle \} , \label{qpset}
\end{array}
\end{equation}
where $\left|0 \right>$ is the qp vacuum and $a^\dagger_{\nu}$ and
$a^\dagger_{\pi}$ the qp creation operators.  The index $\nu$
($\pi$) denotes the neutron (proton) Nilsson quantum numbers, which
run over the orbitals close to the Fermi levels.

Similar to the HFB theory in Ref.\ \cite{Ideguchi10}, we employ a
quadrupole plus pairing Hamiltonian that includes monopole and
quadrupole pairing terms,
\begin{equation}
\hat H = \hat H_0 - {1 \over 2} \chi \sum_\mu \hat Q^\dagger_\mu
\hat Q^{}_\mu - \mpole \hat P^\dagger \hat P - \qpole \sum_\mu \hat
P^\dagger_\mu\hat P^{}_\mu . \label{hamham}
\end{equation}
In Eq.\ (\ref{hamham}), $\hat H_0$ is the spherical single-particle
Hamiltonian, which contains a proper spin-orbit force \cite{Nilsson}.
The monopole pairing strength is taken to be $\mpole = 18.65/A$ and
the quadrupole pairing strength is taken to be $\qpole = 0.2 \mpole$. 

%===============  fig. 1  =============================================
\begin{figure}
\includegraphics[totalheight=7.5cm]{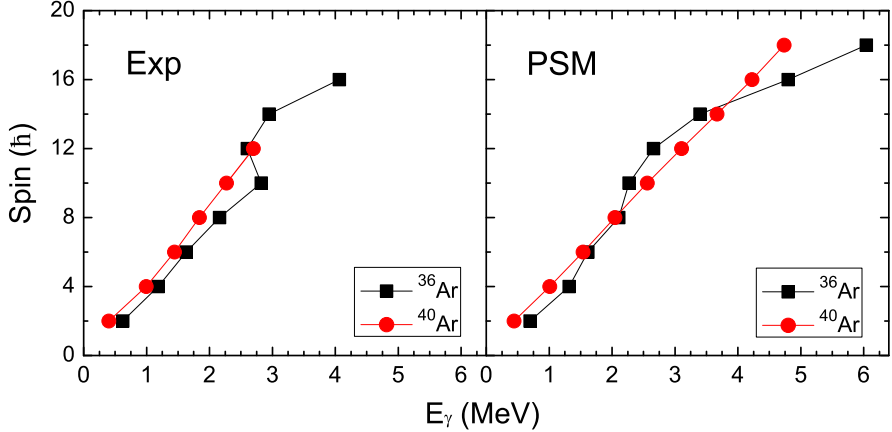}
\caption{(Color online) Calculated $\gamma$-ray energies $E_\gamma =
E(I)-E(I-2)$ (PSM) versus spin $I$ for the superdeformed yrast band in
$^{36,40}$Ar.  These may be  compared with corresponding experimental data (Exp)
taken from Refs.\ \cite{Ar36a} ($^{36}$Ar) and \cite{Ideguchi10}
($^{40}$Ar).} \label{fig1}
\end{figure}
%======================================================================

The PSM calculation for the SD band in $^{40}$Ar is presented in the form of a 
`backbending plot' and compared with experimental data \cite{Ideguchi10} in 
Fig.\ \ref{fig1}.  To emphasize the differences in  rotational behavior 
between $^{40}$Ar and $^{36}$Ar, we  include also the previous PSM results for 
$^{36}$Ar taken from Ref.\ \cite{Long01}. The slope of the curves denotes the 
kinematic moment of inertia, a characteristic quantity for the description of 
rotational behavior. It can be seen that at low spins the behavior of $^{40}$Ar 
is similar to that of $^{36}$Ar. However, in the spin range around $I=12$ the 
two nuclei behave differently.  For $^{40}$Ar, both experiment and calculation 
show approximately a linear relation between the rotational frequency 
($\gamma$-ray energy) and angular momentum, indicating that the superdeformed 
$^{40}$Ar has nearly a constant moment of inertia throughout the entire spin 
range.  In contrast, a band disturbance around $I=12$ is apparent in $^{36}$Ar, 
implying a varying moment of inertia.  

Various model calculations 
\cite{Ar36a,Long01,Inakura02,Oi07,Bender03,Caurier05,Caurier07} interpreted the 
backbending in moment of inertia for the $N=Z$ nucleus $^{36}$Ar as being due to 
a simultaneous alignment of protons and neutrons in the $f_{7/2}$ orbital.  The 
different rotational behavior for the two isotopes was attributed  to different 
occupation of the neutron $f_{7/2}$ orbit in Ref.\ \cite{Ideguchi10}.  With four 
more neutrons in $^{40}$Ar than in $^{36}$Ar,  neutron occupation of the $K=3/2$ 
and 5/2 levels of $f_{7/2}$ in $^{40}$Ar but not in $^{36}$Ar is the main 
source of this difference.

%===============  fig. 2  ============================================
\begin{figure}
\includegraphics[totalheight=10cm]{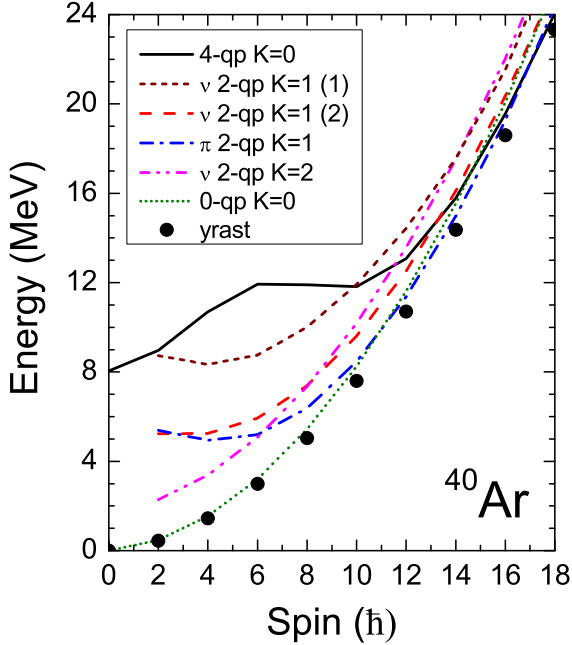}
\caption{(Color online) Band diagram for the superdeformed nucleus
$^{40}$Ar. Only the important lowest-lying bands in each
configuration are shown.} \label{fig2}
\end{figure}
%=====================================================================

The most striking feature of the $^{40}$Ar SD band is the nearly linear 
dependence of $E_\gamma$ on rotation.  For this spin range, rotation alignment 
of particles from particular orbits can occur;  this usually results in a clear 
disturbance of regular rotational sequences, as is seen in $^{36}$Ar.  However, 
this seems not to happen in $^{40}$Ar up to the highest spin measured or 
calculated. To understand this, we have studied the band structures by using 
the band diagram shown in Fig.\ \ref{fig2}.  At deformation 
$\varepsilon_2=0.49$ single-particle orbitals near the neutron and proton Fermi 
levels of $^{40}$Ar are: $K={1\over 2}$, ${3\over 2}$, and ${5\over 2}$ of 
$1f_{7/2}$, $K={1\over 2}$ and ${3\over 2}$ of $1d_{3/2}$, and $K={1\over 2}$ of 
$2p_{3/2}$.  Thus, multi-qp configurations based on combinations of these 
orbitals lie low in energy.  To illustrate their rotational behavior, six 
representative bands are displayed in Fig.\ \ref{fig2}: one 0-qp band (ground 
band), three neutron 2-qp bands, one proton 2-qp band, and one neutron--proton 
4-qp band.  The dots marked  ``yrast" represent the lowest-energy states 
obtained at each spin after band mixing, which were used in Fig.\ 
\ref{fig1} to compare with the data.

The neutron $K=2$ 2-qp band with the configuration $\nu1/2[200] \otimes \nu 
3/2[202]$ originating from $\nu d_{3/2}$ starts with a low energy at about 2.2 
MeV.  However, its rotational behavior is very similar to the g-band.  As can be 
seen in Fig.\ \ref{fig2}, for the whole spin range this band stays nearly 
parallel to the g-band and hence is physically unimportant for yrast 
structure changes.  On the other hand, there are two other 2-qp bands in Fig.\ 
\ref{fig2} beginning at much higher excitation energies ($\sim$5.4 MeV).  The 
proton 2-qp one (labeled $\pi$ 2-qp $K=1$ in Fig.\ \ref{fig2}) with the 
configuration $\pi1/2[330] \otimes \pi 3/2[321]$ shows a unique rotational 
behavior.  As spin increases, the energy initially decreases but then begins 
increasing around spin $I\approx6$.  This behavior has its origin in the spin 
alignment of a decoupled band, as discussed in Ref.\ \cite{PSM}. Because of 
this, it can cross the g-band at $I=12$ and become  the most 
important configuration over the spin interval $I=12-16$.  The other neutron 
2-qp band (labeled $\nu$ 2-qp $K=1(2)$ in Fig.\ \ref{fig2}), with a 
configuration $\nu3/2[321] \otimes \nu 5/2[312]$ coming from $1f_{7/2}$, lies 
higher by several hundred keV and therefore influences the yrast states much 
less. Note that the neutron 2-qp band (marked as $\nu$ 2-qp $K=1(1)$ in Fig.\ 
\ref{fig2}), with the configuration $\nu1/2[330] \otimes \nu 3/2[321]$ from 
$1f_{7/2}$, is found to be roughly 4 MeV above the yrast band. In  $^{36}$Ar, 
this neutron configuration nearly coincides with the same configuration for 
protons and both cross the g-band at the same spin, leading to a simultaneous 
alignment of neutrons and protons \cite{Long01}.  Thus it is clear that the 
addition of four neutrons in $^{40}$Ar shifts the neutron Fermi level up 
relative to that for $^{36}$Ar, which results in a much higher-lying 
configuration $\nu$ 2-qp $K=1(1)$ in $^{40}$Ar. This is the physical reason for 
the difference in rotational behavior for the $^{40}$Ar and $^{36}$Ar SD yrast 
bands.

The 4-qp $K=0$ band (black solid line in Fig.\ \ref{fig2}) consists of the 
two 2-qp $K=1$ bands mentioned above ($\pi$ 2-qp $K=1$ and $\nu$ 2-qp 
$K=1(2)$). It crosses the 2-qp bands at high spins ($I\approx 18$) 
and becomes the lowest band in energy thereafter.  The crossing 
between this 4-qp band and the 2-qp bands is gentle with a very small crossing 
angle. In fact, in the high-spin region after $I=10$ all the low-lying bands 
in Fig.\ \ref{fig2} (0-qp g-band, 2-qp and 4-qp bands) are seen to be bunched 
together and they rotate with the same frequency (given by the slope of the 
curves in Fig.\ \ref{fig2}). Therefore, their successive mutual band crossings  
do not perturb the yrast-band rotation significantly.  This explains the 
nearly-linear dependence of the transition energy $E_\gamma$ on angular momentum 
as seen in Fig.\ \ref{fig1}, which is in contrast to that in the $^{36}$Ar case. 
 Although no disturbance in the yrast energy levels is evident for $^{40}$Ar, 
the corresponding yrast wave functions at high spins are mixed with the 
multi-qp configurations, which may lead to observable effects in other 
quantities such as moments of inertia and electromagnetic properties; we shall 
discuss such effects below.

%===============  fig. 3  =============================================
\begin{figure}
\includegraphics[totalheight=8cm]{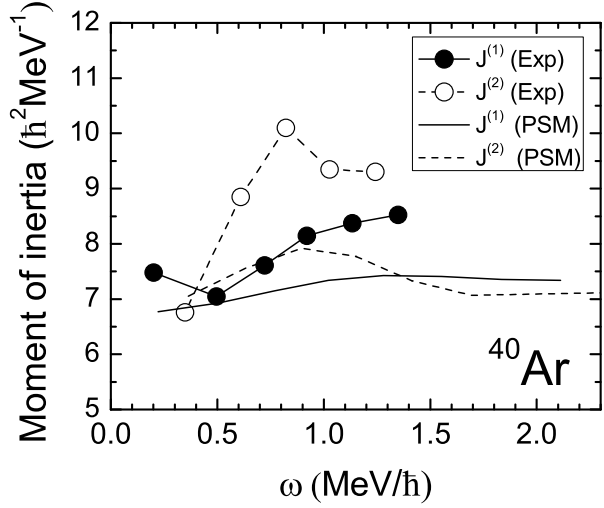}
\caption{Kinematical $(J^{(1)})$ and dynamical $(J^{(2)})$ moments
of inertia of $^{40}$Ar as a function of rotational frequency
$\omega$. Data (Exp)are taken from Ref.\ \cite{Ideguchi10}; PSM denotes 
calculations from this paper. }\label{fig3}
\end{figure}

Experimental \cite{Ideguchi10} and calculated PSM kinematical moments of inertia 
$J^{(1)} (I)$ = $(2I-1)/E_\gamma(I)$ and dynamical moments of inertia  $J^{(2)} 
(I)$ = $4/[E_\gamma(I)-E_\gamma(I-2)]$ for $^{40}$Ar are shown in Fig.\ 
\ref{fig3} as functions of rotational frequency $\omega = E_\gamma/2$, where 
$E_\gamma$ is the transition energy. The experimental $J^{(1)}$ increases with 
rotation while the more sensitive quantity $J^{(2)}$ shows first a more rapid 
climb and then a sudden drop at $\hbar\omega \ge$ 1.0 MeV corresponding to spin 
$I \ge 10\hbar$.  The peak in $J^{(2)}$ suggests a structure change along the SD 
yrast band. These features are qualitatively reproduced by the PSM; it 
nevertheless underestimates the changes.  The drop in $J^{(2)}$ was interpreted 
in Ref.\ \cite{Ideguchi10} as occurring because of the disappearance of the 
pairing gap energy.  In our model, it is due to the first band-crossing of the 
proton 2-qp band with the g-band, as discussed in Fig.\ \ref{fig2}.  Physically, 
it corresponds to the rotation alignment of a pair of $f_{7/2}$ protons.  It is 
important to distinguish the situation found here with the simultaneous 
alignment of both proton and neutrons pairs found in $^{36}$Ar, which causes a 
backbending in the moment of inertia \cite{Ar36a}.

The preceding conclusion for dominance of proton alignment may be examined 
experimentally using $g$-factor measurements.    In the PSM, $g$-factors can be 
computed directly  as
\begin{equation}
g(I) = \frac {\mu(I)}{\mu_N I} = \frac {1}{\mu_N I} \left[ \mu_\pi
(I) + \mu_\nu (I) \right], \label{gfactor}
\end{equation}
with $\mu (I)$ being the magnetic moment of a state $\Psi^I$, and
\begin{eqnarray}
\mu_\tau (I) &=& \left< \Psi^I_{I} | \hat \mu^\tau_z | \Psi^I_{I}
\right> = {I\over{\sqrt{I(I+1)}}} \left<
\Psi^{I} || \hat \mu^\tau || \Psi^{I} \right> \nonumber\\
&=& \frac{I}{\sqrt{I(I+1)}} \left[
     g^{\tau}_l \langle \Psi^I || \hat j^\tau ||\Psi^I
     \rangle + (g^{\tau}_s - g^{\tau}_l)
     \langle \Psi^I || \hat s^\tau || \Psi^I \rangle \right]
     ,
\label{moment}
\end{eqnarray}
where $\tau = \pi$ and $\nu$ for protons and neutrons, respectively.
The following standard values for $g_l$ and $g_s$ appearing in Eq.
(\ref{moment}) are taken:
\begin{eqnarray*}
g_l^\pi &=& 1, ~~~~ g_s^\pi = 5.586 \times 0.75 ,\\
g_l^\nu &=& 0, ~~~~ g_s^\nu = -3.826 \times 0.75 ,
\end{eqnarray*}
and $g_s^\pi$ and $g_s^\nu$ are damped by a usual 0.75 factor from the 
free-nucleon values. Because of the intrinsically opposite signs of the neutron 
and proton $g_s$,  variation of $g$-factors often 
is a clear indicator for a single-particle component that strongly influences 
the total wave function.

%
%
%===============  fig. 4  =============================================
\begin{figure}
\includegraphics[totalheight=8cm]{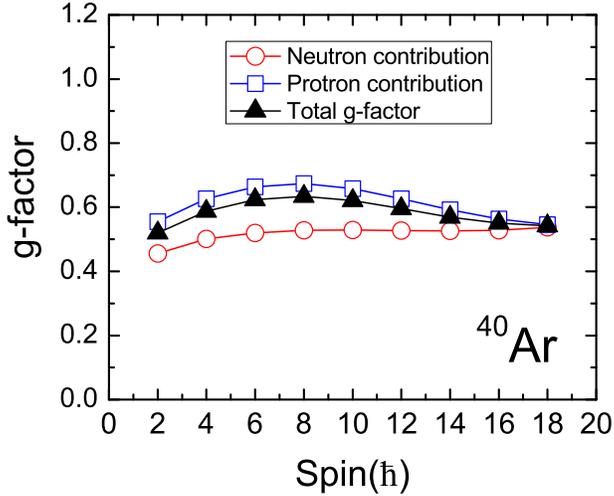}
\caption{Calculated $g$-factors using the PSM for the $^{40}$Ar yrast band. 
}\label{fig4}
\end{figure}

The calculated total $g$-factors displayed in Fig.\ \ref{fig4} (filled 
triangles) suggest non-constant $g$-factors as spin changes, with largest 
values appearing around $I=8$.  An increase in $g$-factor may be attributed to 
the increased proton component in the wave functions. To see this more clearly, 
we plot also in Fig.\ \ref{fig4} curves for the separate contributions 
to the total $g$-factor from neutron (open circles) and proton (open squares) 
qp's.  This is done by eliminating the proton (neutron) qp states in 
(\ref{qpset}) in the calculation for neutron (proton) contribution.  It is clear 
that the shape of the total $g$-factor curve is dominated by the contributions 
from proton qp states.    This reinforces our conclusion that the variation in 
$J^{(2)}$ is caused by a breaking and spin alignment of the $1 f_{7/2}$ proton 
pairs, which contribute the proton components to the total wave functions.  
$g$-factors of the $2^+$ state at 1461 keV in $^{40}$Ar and those in some nearby 
nuclei have been measured \cite{g-factor1,g-factor2,g-factor3}.  We hope that 
the predicted $g$-factors for the SD states in $^{40}$Ar can be verified in the 
future.

%===============  fig. 5  =============================================
\begin{figure}
\includegraphics[totalheight=7cm]{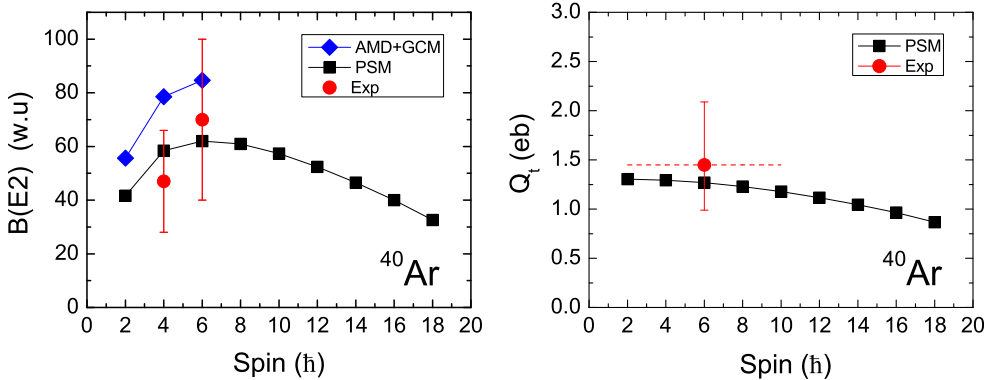}
\caption{(Color online) Left: $B(E2,I\rightarrow I-2)$ values (in W.u.)
for the SD band of $^{40}$Ar compared with available data. Data with
error bars are taken from Ref.\ \cite{BE2DATA04}. Right: Transitional 
quadrupole moments $\qt$ for the SD band of $^{40}$Ar compared with available 
data. Data are taken from Ref.\ \cite{Ideguchi10}, and the horizontal dashed
line indicates that the measured number is an average value for the
spin interval $I=4-12$.} \label{fig5}
\end{figure}

Electric quadrupole transition probabilities $B(E2, I\rightarrow I-2)$ or 
transition quadrupole moment $ \qt (I) $ depend strongly on nuclear shapes.  
They are related by
\begin{equation}
\qt(I) = \sqrt{\frac{16 \pi}{5}} \frac{ \sqrt{B(E2,I\rightarrow
I-2)}} {\langle I, 0, 2, 0 | I-2, 0\rangle } ,
\end{equation}
where $\langle I, 0, 2, 0 | I-2, 0\rangle$ is a Clebsch--Gordan coefficient and 
the $B(E2)$ value for a transition from an initial state of angular momentum $I$ 
to a final state having $I-2$ is given by
\begin{equation}
B(E2, I\rightarrow I-2) = \frac {1}{2I + 1} \left| \right<
\Psi^{I-2} || \hat Q_2 ||\Psi^I \left> \right|^2, \label{BE2}
\end{equation}
with the wavefunctions $\left|\Psi^I\right>$ being those in Eq. (\ref{wf}).  The 
effective charges used in the calculation are the standard ones $e^\pi=1.5e$ and 
$e^\nu=0.5e$.  

The calculated spin-dependent $B(E2)$ and $\qt$ values for the SD band in 
$^{40}$Ar are shown in Fig.\ \ref{fig5}.  The known experimental $B(E2)$ values 
for the $6^+ \rightarrow 4^+$ and $4^+ \rightarrow 2^+$ 
transitions \cite{BE2DATA04} are plotted for comparison.  It can be seen from 
the left plot of Fig.\ \ref{fig5} that the PSM calculation reproduces the 
experimental data nicely, though experimental uncertainties are large.  Our 
calculation predicts the unmeasured low-spin $2^+ \rightarrow 0^+$ and high-spin 
transition probabilities, which may be tested by future experiments.  On the 
other hand, the $B(E2)$ values calculated with the AMD+GCM \cite{Triaxial} 
approach have the right trend but are too large relative to the data; in 
particular the AMD+GCM $4^+ \rightarrow 2^+$ $B(E2)$ is well out of the range of 
the data's error bars.  We found that the band-crossings do not cause drastic 
changes in the PSM $B(E2)$ values, consistent with the finding that the 
transition energies $E_\gamma$ are a linear function of spin.  However, the 
PSM calculation shows that $B(E2)$ values decrease smoothly after spin 
$I=6$, with the $B(E2)$ at $I=18$ being only about 2/3 of that at $I=6$.  The 
PSM calculations thus suggest a decreasing electric quadrupole collectivity 
with increased spin that is caused by successive pair breaking and alignment of 
particles, but considerable collectivity remains still at the highest 
spin states calculated.

As shown in the right plot of Fig.\ \ref{fig5}, our PSM calculation
also predicts large transitional quadrupole moments,
consistent with the experimental data.  The measured value
$\qt=1.45^{+0.49}_{-0.31}$ eb \cite{Ideguchi10} is effectively averaged over
the spin interval $I=4-12$.  As for the $B(E2)$ values, the
calculated $\qt$'s show a smoothly decreasing trend towards high
spins. Let us now consider whether the reduction in $\qt$
or $B(E2)$ with increasing angular momentum is related to triaxiality in the SD 
band.

Taniguchi {\it et al.} \cite{Triaxial} proposed a non-axial superdeformation 
picture and concluded that triaxiality is significant for understanding the 
low-lying states of the SD band in $^{40}$Ar.  This contradicts the conclusion 
of Ideguchi {\it et al.} \cite{Ideguchi10} that axial deformations are 
sufficient to understand these states.  A model comparison for energy levels is 
generally not sufficient to identify triaxiality and one requires more detailed 
information. In Ref.\ \cite{Ring82}, Ring {\it et al.} suggested using two 
measurable quantities, electric quadrupole transition probabilities $B(E2)$ and 
spectroscopic moments $Q$, to extract the two intrinsic deformations $\beta$ 
and $\gamma$ using the relations
\begin{eqnarray}\label{gamma1}
{{\sqrt{B(E2,I\rightarrow I-2)}} \over {\sqrt{B(E2,2\rightarrow
0)}}} &=& {2\over {\sqrt 3}}{\beta^\prime}
\cos(30^\circ-\gamma), \nonumber \\
{{Q(I)} \over {Q(2)}} &=& 2{\beta^\prime} \sin(30^\circ-\gamma),
\end{eqnarray}
where $\beta^\prime$ is the ratio of the
(spin-dependent) $\beta$ deformation parameter to the corresponding
value for a symmetric rotor and the spectroscopic moment is defined through the 
quadrupole matrix element
\begin{equation}
Q(I) = \sqrt{\frac {16\pi}{5}} \left< \Psi^{I}_I | \hat Q_{20}
|\Psi^I_I \right>. \label{Q}
\end{equation}

The model basis for the PSM is axially-symmetric but configuration mixing among 
$K\ne 0$ multi-qp states can generate triaxiality.  Thus  deviation from exact 
axial symmetry should be reflected in the wave functions for PSM solutions, 
leading to a non-zero $\gamma$ extracted from Eqs.\ (\ref{gamma1})--(\ref{Q}). 
To use these equations to determine $\gamma$ deformation parameters we have 
calculated $Q(I)$ in addition to $B(E2,I\rightarrow I-2)$ using the PSM.
\begin{table*}
\caption{Extracted triaxial deformation parameters $\gamma$ (in
degree).}
\begin{tabular}{c|cccccccc}
\hline Spin $I$ & 4 & 6 & 8 & 10 & 12 & 14 & 16 & 18
\\ [1mm]\hline
$\gamma (^\circ)$ & 4.4 & 5.6 & 5.9 & 5.7 & 5.3 & 4.7 & 3.7 & 2.4 \\
\hline
\end{tabular}
{\normalsize } \label{tabel1}
\end{table*}
%
%
%===============  fig. 6  =============================================
\begin{figure}
\includegraphics[totalheight=8cm]{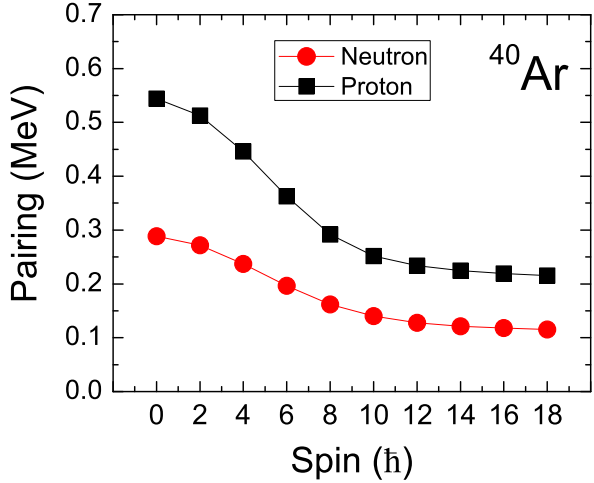}
\caption{(Color online) Calculated pairing gaps for the PSM.} \label{fig6}
\end{figure}
The values of the spin-dependent triaxiality parameter $\gamma$ 
obtained from this analysis are listed in Table I.  One sees that non-zero 
triaxiality is indeed found in our calculation; however, the deduced $\gamma$ 
values are very small.  

In Ref.\ \cite{Oi07}, Oi pointed out that in the CHFB calculation pairing 
correlations tend to suppress triaxial deformation.  Without pairing, triaxial 
deformation would be enhanced and ultimately emerge at high spins.  Pairing 
correlation is taken into account explicitly in the PSM.  In Fig.\ 
\ref{fig6}, we show the calculated pairing gaps, computed using expectation 
values of the pair operator with respect to the PSM wave functions.  It is found 
that for $^{40}$Ar, neutron and proton pairing gaps are not small; for example 
$\Delta_n=0.55$ 1 MeV at $I=0$.  However, the pairing gaps decrease as the 
nucleus rotates, saturating eventually at smaller but non-zero values for high 
spins.  For the high-spin states, the remaining pairing correlation is 
relatively weak but still may play a role in sustaining collectivity so that the 
triaxiality is suppressed to $|\gamma|<10^\circ$, as suggested in Ref.\ 
\cite{Oi07}.  Thus we conclude that $^{40}$Ar has a nearly axially-symmetric 
shape and triaxiality is not very important for the SD states in this nucleus.

To summarize, experimental data for the SD rotational band in $^{40}$Ar 
\cite{Ideguchi10} have provided a valuable example that allows studying the 
stabilization of a superdeformed minimum in light nuclei. For this nucleus, 
there has been a debate about whether it has an axially or triaxially deformed 
shape \cite{Triaxial}.  The present study has applied the projected shell model 
that was used before in the calculation of SD bands in this mass region 
\cite{Long01}.  Our results have shown a good reproduction of the newer 
experimental data and explained the rotational behavior of the SD band as 
successive band crossings of the 0-, 2-, and 4-qp configurations.  Calculated 
$g$-factors have been presented to support the importance of proton alignment 
in the wave functions.  The calculated electric quadrupole transition 
probabilities reproduce well the known experimental data, and suggest a 
reduced but still significant collectivity in the high spin-region.  Triaxial 
deformation parameters have been deduced from the calculated $B(E2)$ and  
spectroscopic quadrupole moment $Q$, both of which are experimentally 
measurable. The triaxial deformation parameter $\gamma$ has been found to 
average only a few degrees throughout the SD band, suggesting that triaxiality 
is unlikely to be an important factor in understanding superdeformation in 
$^{40}$Ar.

%======================================================================

Research at SJTU was supported by the National Natural Science
Foundation of China (No.\ 11135005), by the 973 Program of China (No.\
2013CB834401), and by the Open Project Program of the State Key
Laboratory of Theoretical Physics, Institute of Theoretical Physics,
Chinese Academy of Sciences, China (No.\ Y5KF141CJ1).


\begin{thebibliography}{99}

\bibitem{Frauendorf01}
S. Frauendorf, Rev.\ Mod.\ Phys.\ {\bf 73} (2001) 463.

\bibitem{SD-review}
P. J. Nolan and P. J. Twin, Ann.\ Rev.\ Nucl.\ Part.\ Sci., {\bf 38}
(1998) 533.

\bibitem{SD-data}
X.-L. Han and C.-L. Wu, At.\ Data Nucl.\ Data Tables {\bf 73} (1999)
43.

\bibitem{SD190}
Y. Sun, J.-y.\ Zhang, M. Guidry, Phys.\ Rev.\ Lett.\ {\bf 78} (1997)
2321.

\bibitem{SD60}
Y. Sun, J.-y.\ Zhang, M. Guidry, C.-L. Wu, Phys.\ Rev.\ Lett.\ {\bf 83}
(1999) 686.

\bibitem{Ideguchi10}
E. Ideguchi, {\it et al.}, Phys. Lett. B {\bf 686} (2010) 18.

\bibitem{Ar36a}
C. E. Svensson, {\it et al.}, Phys.\ Rev.\ Lett.\ {\bf 85} (2000) 2693.

\bibitem {Ar36b}
C. E. Svensson, {\it et al.}, Phys.\ Rev.\ C {\bf 63} (2001)
061301(R).

\bibitem{Long01}
G.-L. Long and Y. Sun, Phys.\ Rev.\ C {\bf 63} (2001) 021305(R).

\bibitem{BE2DATA04}
J. A. Cameron and B. Singh, Nucl.\ Data Sheets {\bf 102} (2004) 293.

% N=Z light Nuclei of 36Ar, 40Ca and 44Ti
\bibitem{Ideguchi01}
E. Ideguchi, {\it et al.}, Phys.\ Rev.\ Lett.\ {\bf 87} (2001) 222501.

\bibitem{Leary00}
C. D. O`Leary, {\it et al.}, Phys.\ Rev.\ C {\bf 61} (2000) 064314.

\bibitem{Mg34}
H. Iwasaki, {\it et al.}, Phys.\ Lett.\ B {\bf 522} (2001) 227.

\bibitem{Brown90}
E. K. Warburton, J. A. Becker, B. A. Brown, Phys.\ Rev.\ C {\bf 41}
(1990) 1147.

\bibitem{Triaxial}
Y. Taniguchi, {\it et al.}, Phys.\ Rev.\ C {\bf 82} (2010) 011302(R).

\bibitem{Rubbia98}
A. Rubbia, Nucl.\ Phys.\ B {\bf 66} (1998) 436.

\bibitem{Niu14}
B.-N. Niu, E. Hiyama, H. Sagawa, and S.-G. Zhou, Phys.\ Rev.\ C {\bf
89} (2014) 044307.

\bibitem{Inakura02}
T. Inakura, {\it et al.}, Nucl.\ Phys.\ A {\bf 710} (2002) 261.

\bibitem{Poves04} A. Poves, Nucl.\ Phys.\ A {\bf 731} (2004) 339.

\bibitem{Oi07}
M. Oi, Phys.\ Rev.\ C {\bf 76} (2007) 044308.

\bibitem{Bender03}
M. Bender, H. Flocard, P.-H. Heenen, Phys.\ Rev.\ C {\bf 68} (2003)
044321.

%\bibitem{Caurier94}
%E. Caurier, {\it et al.}, Phys. Rev. C {\bf 50} (1994) 225.

\bibitem{PSM} K. Hara, Y. Sun, Int.\ J.\ Mod.\ Phys.\ E {\bf 4} (1995) 637.

\bibitem{SD130} Y. Sun and M. Guidry, Phys.\ Rev.\ {\bf C52} (1995) R2844.

\bibitem{Nilsson}
T. Bengtsson and I. Ragnarsson, Nucl.\ Phys.\ A {\bf 436} (1985) 14.

\bibitem{Caurier05}
E. Caurier, {\it et al.}, Phys.\ Rev.\ Lett.\ {\bf 95} (2005) 042502.

\bibitem{Caurier07}
E. Caurier, {\it et al.}, Phys.\ Rev.\ C {\bf 75} (2007) 054317.

\bibitem{Cr48} K. Hara, Y. Sun, T. Mizusaki,
Phys.\ Rev.\ Lett.\ {\bf 83} (1999) 1922.

%\bibitem{Mg32}
%T. Motobayashi, {\it et al.}, Phys. Lett. B {\bf 346} (1995) 9.

% Ar40 g-factor
\bibitem{g-factor1}
E. A. Stefanova, {\it et al.}, Phys.\ Rev.\ C {\bf 72} (2005) 014309.

\bibitem{g-factor2}
A. E. Stuchbery, {\it et al.}, Phys.\ Rev.\ C {\bf 74} (2006) 054307.

\bibitem{g-factor3}
K. H. Speidel, {\it et al.}, Phys.\ Rev.\ C {\bf 78} (2008) 017304.

\bibitem{Ring82}
P. Ring, A. Hayashi, K. Hara, H. Emling, E. Grosse, Phys.\ Lett.\ {\bf
110B} (1982) 423.

\end{thebibliography}
\end{document}